\documentclass[prc,reprint,superscriptaddress,showpacs,showkeys]{revtex4-1}

\usepackage[english]{babel}
\usepackage[utf8]{inputenc}
\usepackage[T1]{fontenc}
\usepackage{amsfonts}
\usepackage{amssymb}
\usepackage{amsmath}
\usepackage{bm}
\usepackage{dsfont}
\usepackage{braket}
\usepackage{simplewick}
\usepackage{slashed}
\usepackage[version=4]{mhchem}
\usepackage{array}
\usepackage{dcolumn}
\usepackage{longtable}
\usepackage{rotating}
\usepackage{graphicx}
\usepackage{xcolor}
\usepackage[colorlinks]{hyperref}

\newcolumntype{d}[1]{D{.}{.}{#1}}

\definecolor{color_01}{rgb}{0,0,0.75}
\definecolor{color_02}{rgb}{0.75,0,0}
\definecolor{color_03}{rgb}{0.5,0,0.75}
\hypersetup{linktoc=page,linkcolor=color_01,citecolor=color_02,urlcolor=color_03}

\begin{document}
\title{Bound-state double-$\beta$ decay}

\author{A.~Babi\v{c}}
\affiliation{Faculty of Nuclear Sciences and Physical Engineering, Czech Technical University in Prague, 115~19~Prague, Czech Republic}
\affiliation{Institute of Experimental and Applied Physics, Czech Technical University in Prague, 128~00~Prague, Czech Republic}
\affiliation{Bogoliubov Laboratory of Theoretical Physics, Joint Institute for Nuclear Research, 141980~Dubna, Russia}
\author{D.~\v{S}tef\'{a}nik}
\affiliation{Faculty of Mathematics, Physics and Informatics, Comenius University in Bratislava, 842~48~Bratislava, Slovakia}
\author{M.~I.~Krivoruchenko}
\affiliation{Bogoliubov Laboratory of Theoretical Physics, Joint Institute for Nuclear Research, 141980~Dubna, Russia}
\affiliation{Institute for Theoretical and Experimental Physics, 117259~Moscow, Russia}
\author{F.~\v{S}imkovic}
\affiliation{Institute of Experimental and Applied Physics, Czech Technical University in Prague, 128~00~Prague, Czech Republic}
\affiliation{Bogoliubov Laboratory of Theoretical Physics, Joint Institute for Nuclear Research, 141980~Dubna, Russia}
\affiliation{Faculty of Mathematics, Physics and Informatics, Comenius University in Bratislava, 842~48~Bratislava, Slovakia}

\date{\today}

\begin{abstract}
We consider new modes of two-neutrino and neutrinoless double-$\beta$ decays in which one $\beta$ electron goes over to a continuous spectrum and the other occupies a vacant bound level of the daughter ion. We calculate the corresponding phase-space factors of the final states, estimate the partial decay rates, and derive the one- and two-electron energy spectra using relativistic many-electron wave functions of atoms provided by the multiconfiguration Dirac--Hartree--Fock package \textsc{Grasp2K}. While the bound-state neutrinoless double-$\beta$ decays are strongly suppressed, their two-neutrino counterparts can be observed in the next-generation double-$\beta$-decay experiments, most notably SuperNEMO.
\end{abstract}


\maketitle
\section{Introduction}
Among the most challenging problems of modern neutrino physics are the mechanism of neutrino mixing and the nature of neutrino masses (Dirac or Majorana). If diagonal neutrinos $\nu_i$ ($i = 1, \, 2, \, 3$) are Majorana fermions, then flavor neutrinos $\nu_{\alpha}$ ($\alpha = e, \, \mu, \, \tau$) are identical to their charge-conjugated states, as a result of which the total lepton number is not conserved (see, e.g., \cite{Bil87}). Observation of the neutrinoless double-$\beta$ ($0 \nu \beta \beta$) decay can provide evidence for the Majorana nature of massive neutrinos, which would be of great value for extensions of the Standard Model \cite{Ver16}. Measurement of the half-life of the $0 \nu \beta \beta$ decay could provide a key to the absolute scale of neutrino masses and also shed light on the leptonic CP violation mechanism required to explain the observed baryon asymmetry of the Universe \cite{Ver16,Bab13}. Given the opportunity to get answers to so many fundamental questions, the $0 \nu \beta \beta$ decay has attracted much attention of theorists and experimentalists in the recent decades.

The neutrinoless (two-neutrino) double-$\beta$ decay of a parent nucleus $\ce{_Z^AX}$ into a daughter nucleus $\ce{_{Z + 2}^AY}$, denoted $0 \nu (2 \nu) \beta \beta$, involves the emission of two electrons $e^-$ (and a pair of electron antineutrinos $\overline{\nu}_e$) from the atom:
\begin{equation}
\ce{_Z^AX} \longrightarrow \ce{_{Z + 2}^AY} + e^- + e^- + (\overline{\nu}_e + \overline{\nu}_e).
\end{equation}
The $2 \nu \beta \beta$ decay occurs in the 2nd order of weak interaction and as such it conserves the total lepton number: $\Delta L = 0$. It forms the dominant decay channel of beta radioactivity of even-even isotopes for which the single-$\beta$ decay into the odd-odd intermediate nucleus is either energetically forbidden or suppressed by spin selection rules. The double-$\beta$ decay has so far been observed in 11 out of 35 candidate isotopes, with half-lives $T_{1/2}^{2 \nu \beta \beta} \sim 10^{19} \textrm{--} 10^{21} \, \mathrm{yr}$, making it the rarest known spontaneous decay in nuclear physics. In contrast, the $0 \nu \beta \beta$ decay violates the total lepton number by two units: $\Delta L = +2$, and requires a Majorana mass term. This process could be observed as a monoenergetic peak at the $2 \nu \beta \beta$ spectrum endpoint in calorimetric measurements of the sum of electron energies. The current limits on the half-lives set $T_{1/2}^{0 \nu \beta \beta} > (0.18 \textrm{--} 1.07) \times 10^{26} \, \mathrm{yr}$ at $90\% \, \mathrm{C.L.}$ for the $\ce{^{136}Xe}$ and $\ce{^{76}Ge}$ isotopes \cite{Alb18,Ago18,Gan16}.

In 1961, Bahcall \cite{Bah61} developed a formalism for description of bound-state $\beta$ decays in which the $\beta$-electron is produced in an atomic $\mathrm{K}$ or $\mathrm{L}$ shell, while the monochromatic antineutrino $\overline{\nu}_e$ carries away the entire energy of the decay. The bound-state $\beta$ decay was observed on bare $\ce{_{66}^{163}Dy^{66+}}$ ions collected in the heavy-ion storage ring ESR at GSI, Darmstadt, with a half-life of $47 \, \mathrm{d}$ for the otherwise stable nuclide \cite{Jun92}.

The neutrinoless double-$\beta$ decay with two bound electrons in the final state denoted by $0 \nu \mathrm{EP} \mathrm{EP}$ (where $\mathrm{EP}$ stands for the ``electron placement''):
\begin{equation}
\ce{_Z^AX} \longrightarrow \ce{_{Z + 2}^AY^*} + e_{\mathrm{b}}^- + e_{\mathrm{b}}^-
\end{equation}
was discussed in Ref.~\cite{Kri11} as an inverse process to the neutrinoless double-electron capture. Resonant enhancement of the $0 \nu \mathrm{EP} \mathrm{EP}$ decay probability can occur in the case of quasi-degeneracy of the initial- and final-state atomic energies. The ground-state $0^+ \longrightarrow 0^+$ nuclear transition of $\ce{^{148}Nd}$ to an $1.921 \, \mathrm{MeV}$ excited state of $\ce{^{148}Sm^*}$ fulfills the resonance condition with the experimental accuracy of $\approx 10 \, \mathrm{keV}$. The estimated half-life, however, was found to be beyond the reach of experiments at the present stage.

\begin{figure}[t]
\includegraphics[width=0.5\columnwidth]{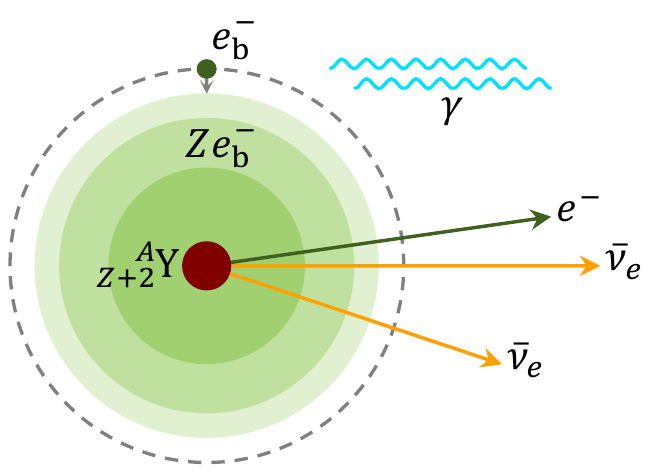}
\caption{\label{fig:view}A schematic view of the $0 \nu (2 \nu) \mathrm{EP} \beta$ decays. The final state involves the daughter nucleus $\ce{_{Z + 2}^AY}$, a bound electron $e_{\mathrm{b}}^-$ produced above the subshells occupied by $Z$ atomic electrons, and a single free electron $e^-$ (and a pair of electron antineutrinos $\overline{\nu}_e$) emitted from the atom. Upon the deexcitation, the bound electron $e_{\mathrm{b}}^-$ radiates photons of energy $\lesssim 10 \, \mathrm{eV}$.}
\end{figure}

In this paper, we develop a formalism for description of the bound-state two-neutrino and neutrinoless double-$\beta$ decays denoted by $0 \nu (2 \nu) \mathrm{EP} \beta$:
\begin{equation}
\ce{_Z^AX} \longrightarrow \ce{_{Z + 2}^AY} + e_{\mathrm{b}}^- + e^- + (\overline{\nu}_e + \overline{\nu}_e).
\end{equation}
The process is shown schematically in Fig.~\ref{fig:view}. The appearance of the first $\beta$-electron in the continuous energy spectrum is accompanied by a production of the second $\beta$-electron in a vacant discrete $n\mathrm{s}_{1/2}$ or $n\mathrm{p}_{1/2}$ level above the valence shell of the daughter ion $\ce{_{Z + 2}^AY^{2+}}$. The inclusion of atomic levels with higher total angular momenta is not required because their wave functions exhibit only a negligible overlap with the nucleus. Since the $0 \nu \mathrm{EP} \beta$, $0 \nu \beta \beta$, $2 \nu \mathrm{EP} \beta$ and $2 \nu \beta \beta$ decay modes constitute 1-, 2-, 3- and 4-body decays, respectively, they could be distinguished by their one- and two-electron energy distributions.

The outline of the paper is as follows. In Sec.~\ref{sec:rewf}, the relativistic electron wave functions as one-particle solutions to the Dirac equation are described and expressions for the relativistic Fermi function and its bound-state analog are derived. In Sec.~\ref{sec:psf}, the $0 \nu (2 \nu) \mathrm{EP} \beta$ decay rates are derived within the $\mathrm{V} - \mathrm{A}$ weak interaction theory including the mixing of Majorana neutrinos. We restrict ourselves to the ground-state $0^+ \longrightarrow 0^+$ nuclear transitions and obtain the phase-space factors entering into the decay rates. Section~\ref{sec:dhf} describes the evaluation of relativistic bound-electron wave functions at short distances via the multiconfiguration Dirac--Hartree--Fock package \textsc{Grasp2K} \cite{Jon13}. Numerical estimates of the half-lives and the $0 \nu (2 \nu) \mathrm{EP} \beta$ to $0 \nu (2 \nu) \beta \beta$ decay-rate ratios are given in Sec.~\ref{sec:rd} in addition to the one- and two-electron energy spectra. In Sec.~\ref{sec:c}, we finally draw conclusions regarding possible experimental observation of the bound-state double-$\beta$ decays and provide motivation for further studies.

\section{Relativistic electron wave functions in central field}
\label{sec:rewf}
The electronic structure of atoms is described by the shell-model relativistic wave functions obtained as solutions to the Dirac equation in a self-consistent centrally symmetric potential which is a superposition of the nuclear Coulomb potential and the screening potential of the electron shell. The corresponding bispinors with separated radial ($r = |\mathbf{r}|$) and angular ($\mathbf{n} = \mathbf{r}/|\mathbf{r}|$) variables take the form (see, e.g., \cite{Ber82}):
\begin{equation}
\psi_{\kappa \mu}(\mathbf{r}) =
\begin{pmatrix}
f_{\kappa}(r) \, \Omega_{\kappa \mu}(\mathbf{n}) \\
i g_{\kappa}(r) \, \Omega_{-\kappa \mu}(\mathbf{n})
\end{pmatrix},
\end{equation}
where $\kappa = (l - j)(2j + 1) = \pm 1, \, \pm 2, \, \dots$ labels combinations of the orbital $l = 0, \, 1, \, \dots$ and spin $s = 1/2$ angular momenta ($\kappa = -1, \, +1$ for $n\mathrm{s}_{1/2}$ and $n\mathrm{p}_{1/2}$ states, respectively), while $\mu = -j, \, \dots \, , \, +j$ denotes the projection of the total angular momentum $\mathbf{j} = \mathbf{l} + \mathbf{s}$ onto the $z$-axis. The spherical spinors with parity $(-1)^l$ are defined by:
\begin{equation}
\Omega_{\kappa \mu}(\mathbf{n}) = \sum_{\sigma = \pm 1/2} C_{l \, \mu - \sigma \, \frac{1}{2} \, \sigma}^{j \mu} \, Y_{l \, \mu - \sigma}(\mathbf{n}) \, \chi_{\sigma},
\end{equation}
where $C_{l \, \mu - \sigma \, \frac{1}{2} \, \sigma}^{j \mu}$ are the Clebsch--Gordan coefficients, $\chi_{\sigma}$ are two-component spinors, and $\sigma$ is the spin projection.

The radial functions $f_{\kappa}(r)$ and $g_{\kappa}(r)$ in the continuum further depend on the electron energy $E$. In the double-$\beta$ decays, the leading $\mathrm{s}_{1/2}$ term of the partial-wave expansion which enters the nuclear matrix elements reads \cite{Doi85}:
\begin{equation}
\psi_{\mathrm{s}_{1/2}}(\mathbf{p}, \, \mathbf{r}) =
\begin{pmatrix}
f_{-1}(E, \, r) \, \chi \\
g_{+1}(E, \, r) \, \boldsymbol{\sigma} \cdot \hat{\mathbf{p}} \, \chi
\end{pmatrix},
\end{equation}
where $\hat{\mathbf{p}}$ is a unit vector in the direction of the electron momentum $\mathbf{p}$. The continuum radial functions are normalized to the $\delta$ function in $p = |\mathbf{p}|$, while the bound states obey: $\int \mathrm{d}r \, r^2 (f^2 + g^2) = 1$.

The Fermi function $F(Z, \, E)$, introduced to correct the short-distance behavior of the $\beta$-electron plane waves due to the Coulomb potential, is defined in terms of the radial wave functions $f_{-1}(E, \, r)$ and $g_{+1}(E, \, r)$ evaluated at the nuclear surface at $r = R \approx 1.2 \, \mathrm{fm} \, A^{1/3}$:
\begin{align}
\label{eq:f}
F(Z, \, E) & = f_{-1}^2(E, \, R) + g_{+1}^2(E, \, R) \nonumber \\
& \approx 4 \left[ \frac{\left| \Gamma(\gamma + i \nu) \right|}{\Gamma(2 \gamma + 1)} \right]^2 (2pR)^{2 \gamma - 2} \, e^{\pi \nu},
\end{align}
where $\gamma = \sqrt{\kappa^2 - (\alpha Z)^2}$, $\nu = \alpha Z E/p$, $p = \sqrt{E^2 - m_e^2}$, $m_e$ is the electron mass, and $\alpha \approx 1/137$ is the fine-structure constant. We remark that $F(Z, \, E) \to 1$ for $Z \to 0$. For $\alpha Z \ll 1$ and $l = 0$, the Fermi function $F(Z, \, E)$ coincides with the Gamow--Sommerfeld factor \cite{Gam28,Gam29,Som23}.

The Fermi function in Eq.~(\ref{eq:f}) is given by standard approximation \cite{Doi85} in which the relativistic electron wave function for a uniform charge distribution in the nucleus is considered and only the lowest-order terms in the power expansion in $r$ are taken into account. The exact Dirac electron wave function accounting for a finite nuclear size and electron-shell screening effects \cite{Kot12} modifies the $0 \nu \beta \beta$-decay phase-space factor for $\ce{^{150}Nd}$ by $30\%$ (see Ref.~\cite{Ste15} and Table~1 therein), which results in an increase in the $0 \nu \beta \beta$-decay half-life. The $0 \nu (2 \nu) \mathrm{EP} \beta$ decay rate with one electron in the continuous spectrum is thus less sensitive to the details of the Dirac electron wave function since only one Fermi function enters the corresponding phase-space factor. We therefore restrict ourselves to the continuous-spectrum solutions of the Coulomb problem for $V(r) = -\alpha (Z + 2)/r$, where $Z + 2$ is the atomic number of the daughter nucleus $\ce{_{Z + 2}^AY}$.

In the discrete spectrum, the radial wave functions $f_{n \kappa}(r)$ and $g_{n \kappa}(r)$ in the Coulomb potential correspond to the energy eigenvalues (see, e.g., \cite{Ber82}):
\begin{equation}
E_{n \kappa} = m_e \left[ 1 + \frac{(\alpha Z)^2}{(\gamma + n_r)^2} \right]^{-\frac{1}{2}},
\end{equation}
where $n = 1, \, 2, \, \dots$ is the principal quantum number and $n_r = n - |\kappa|$ is the radial quantum number which counts the number of radial nodes. At small distances $r \sim R \ll 1/\lambda$, where $\lambda = \sqrt{m_e^2 - E_{n \kappa}^2}$, the leading term from the series expansion of the radial wave functions $f_{n \kappa}(r)$ and $g_{n \kappa}(r)$ for a point-like source can be found in Ref.~\cite{Kri11}.

The radial wave functions enter the bound-state $\beta$-decay probabilities in the combination:
\begin{equation}
\label{eq:b}
B_n(Z) = f_{n, -1}^2(R) + g_{n, +1}^2(R),
\end{equation}
which formally coincides with the relativistic Fermi function (\ref{eq:f}). Note that the first and second terms in the right-hand side of Eq.~(\ref{eq:b}) originate from the production of $\beta$-electrons in the $n\mathrm{s}_{1/2}$ and $n\mathrm{p}_{1/2}$ orbits, respectively. For $\alpha Z \ll 1$ and $l = 0$, we have $f_{n \kappa}(r) \approx R_{nl}(r)$ and $g_{n \kappa}(r) \approx 0$, where $R_{nl}(r)$ is the nonrelativistic radial wave function obtained by solving the Schr\"{o}dinger wave equation for a hydrogen-like atom. The screening of the Coulomb potential modifies the short-distance behavior of the bound-state wave functions. This effect is taken into account in Sec.~\ref{sec:dhf} via the relativistic atomic structure package \textsc{Grasp2K}.

\section{Phase-space factors}
\label{sec:psf}
The double-$\beta$ decay is the 2nd-order process governed by the effective $\beta$-decay Hamiltonian:
\begin{equation}
\mathcal{H}_{\beta} = \frac{G_{\beta}}{\sqrt{2}} \, \overline{e} \, \gamma^{\mu} \left( 1 - \gamma^5 \right) \nu_e \, j_{\mu} + \mathrm{H.c.}
\end{equation}
Here, $G_{\beta} = G_{\mathrm{F}} \cos \theta_{\mathrm{C}}$ includes the Fermi coupling constant $G_{\mathrm{F}} \approx 1.166 \times 10^{-5} \, \mathrm{GeV}^{-2}$ together with the Cabibbo angle $\theta_{\mathrm{C}} \approx 13^{\circ}$ due to the quark mixing \cite{Pat16}, $e$ and $\nu_e$ are the electron and electron-neutrino fields, respectively, and the baryon charged current $j_{\mu} = \overline{p} \, \gamma_{\mu} \left( g_V - g_A \, \gamma^5 \right) \, n$ couples the proton and neutron fields via the vector $g_V = 1$ and (unquenched) axial-vector $g_A \approx 1.27$ coupling constants. The $\mathrm{V} - \mathrm{A}$ structure of $\mathcal{H}_{\beta}$ ensures that only the left-handed leptons participate in the weak interaction. The flavor- and diagonal-neutrino fields are related by the unitary $3 \times 3$ Pontecorvo--Maki--Nakagawa--Sakata (PMNS) matrix $U$:
\begin{equation}
\nu_{\alpha} = \sum_i U_{\alpha i} \, \nu_i.
\end{equation}
The neutrinoless double-$\beta$ decay is assumed to be related to a light Majorana-neutrino exchange between nucleons in the parent nucleus.

The inverse $0 \nu \beta \beta$ and $2 \nu \beta \beta$ half-lives (see, e.g., \cite{Kot12}):
\begin{align}
\left( T_{1/2}^{0 \nu \beta \beta} \right)^{-1} & = g_A^4 \, G^{0 \nu \beta \beta}(Z, \, Q) \left| M^{0 \nu \beta \beta} \right|^2 \left| \frac{m_{\beta \beta}}{m_e} \right|^2, \nonumber \\
\left( T_{1/2}^{2 \nu \beta \beta} \right)^{-1} & = g_A^4 \, G^{2 \nu \beta \beta}(Z, \, Q) \left| m_e \, M^{2 \nu \beta \beta} \right|^2
\end{align}
factorize in terms of the kinematic phase-space factors $G^{0 \nu (2 \nu) \beta \beta}(Z, \, Q)$, the nuclear matrix elements (NMEs) $M^{0 \nu (2 \nu) \beta \beta}$, and the effective Majorana neutrino mass:
\begin{equation}
m_{\beta \beta} = \sum_i U_{ei}^2 \, m_i,
\end{equation}
where $m_i$ are the masses of diagonal neutrinos. Since the absolute scale of neutrino masses and the Majorana phases are unknown, the value of $\left| m_{\beta \beta} \right|$ is treated as a parameter. The experimental lower bounds on $T_{1/2}^{0 \nu \beta \beta}$ set an upper limit on $\left| m_{\beta \beta} \right|$. The most stringent limit has so far been obtained in the KamLAND-Zen experiment \cite{Gan16}: $\left| m_{\beta \beta} \right| < 61 \textrm{--} 165 \, \mathrm{meV}$ at $90\% \, \mathrm{C.L.}$, where the range of values accounts for the uncertainties inherent in the nuclear-structure models. In the case of the inverted hierarchy of neutrino masses, the effective mass is constrained by cosmology: $\left| m_{\beta \beta} \right| = 20 \textrm{--} 50 \, \mathrm{meV}$. We estimate the $0 \nu \mathrm{EP} \beta$ and $0 \nu \beta \beta$ half-lives assuming $\left| m_{\beta \beta} \right| = 50 \, \mathrm{meV}$. Since the $2 \nu \beta \beta$ half-life is unambiguously defined within the Standard Model, the measured values of $T_{1/2}^{2 \nu \beta \beta}$ can be used to fix the phenomenological parameters, improve the predictions of the nuclear-structure models for $M^{0 \nu \beta \beta}$ and probe the possible quenching of $g_A$.

The energy conservation in the $0 \nu (2 \nu) \beta \beta$ decays implies: $M_i = M_f + E_1 + E_2 + (\omega_1 + \omega_2)$, where $M_i$ and $M_f$ are the masses of the parent and daughter nuclei, and $E_1$ and $E_2$ (and $\omega_1$ and $\omega_2$) are the total energies of the emitted electrons (and antineutrinos), respectively. The total released kinetic energy in both scenarios equals: $Q = M_i - M_f - 2 m_e$. Due to indistinguishability of the final-state leptons, the NMEs contain a superposition of two (four) energy denominators \cite{Bil10}:
\begin{align}
& M^{0 \nu \beta \beta}: \, \frac{1}{E_n - M_i + E_{1, 2} + q^0} \approx \frac{1}{E_n - \frac{M_i + M_f}{2} + q^0}, \nonumber \\
& M^{2 \nu \beta \beta}: \, \frac{1}{E_n - M_i + E_{1, 2} + \omega_{1, 2}} \approx \frac{1}{E_n - \frac{M_i + M_f}{2}},
\end{align}
where $E_n$ denotes the $n$th energy level of the intermediate nucleus and $q = (q^0, \, \mathbf{q})$ is the four-momentum of the exchanged Majorana neutrino. Since $q^0 = \sqrt{\mathbf{q}^2 + m_i^2} \approx |\mathbf{q}| \sim 200 \, \mathrm{MeV}$, the difference between the lepton energies can be safely neglected: $-M_i + E_{1, 2} + (\omega_{1, 2}) = -\frac{M_i + M_f}{2} \pm \frac{E_1 - E_2}{2} \pm \left( \frac{\omega_1 - \omega_2}{2} \right) \approx -\frac{M_i + M_f}{2}$. In case of the $0 \nu (2 \nu) \mathrm{EP} \beta$ decay modes, a similar approximation ensures that the corresponding NMEs remain essentially unchanged: $M^{0 \nu (2 \nu) \mathrm{EP} \beta} \approx M^{0 \nu (2 \nu) \beta \beta}$ and the distinction between the $0 \nu (2 \nu) \mathrm{EP} \beta$ and $0 \nu (2 \nu) \beta \beta$ decay modes is fully captured by the phase-space factors $G^{0 \nu (2 \nu) \mathrm{EP} \beta}(Z, \, Q)$.

The phase-space factors of the $0 \nu (2 \nu) \mathrm{EP} \beta$ decays can be found to be:
\begin{align}
\label{eq:g0n}
G^{0 \nu \mathrm{EP} \beta} & = \frac{G_{\beta}^4 \, m_e^2}{32 \pi^4 R^2 \ln 2} \sum_{n = n_{\mathrm{min}}}^{\infty} B_n(Z + 2) \, F(Z + 2, \, E) \, E \, p, \\
\label{eq:g2n}
G^{2 \nu \mathrm{EP} \beta} & = \frac{G_{\beta}^4}{8 \pi^6 m_e^2 \ln 2} \sum_{n = n_{\mathrm{min}}}^{\infty} B_n(Z + 2) \nonumber \\
& \quad \times \int\limits_{m_e}^{m_e + Q} \mathrm{d}E \, F(Z + 2, \, E) \, E \, p \int\limits_0^{m_e + Q - E} \mathrm{d}\omega_1 \, \omega_1^2 \, \omega_2^2,
\end{align}
where $n_{\mathrm{min}}$ is the principal quantum number of the lowest vacant electron shell (this can in principle be different for the $\mathrm{s}_{1/2}$ and $\mathrm{p}_{1/2}$ states). Equations~(\ref{eq:g0n})--(\ref{eq:g2n}) can be derived from $G^{0 \nu \beta \beta}$ and $G^{2 \nu \beta \beta}$ using the substitution:
\begin{equation}
\label{eq:fb1}
\frac{\mathrm{d} \mathbf{p}}{(2 \pi)^3} \, F(Z + 2, \, E) \longmapsto \frac{1}{4 \pi} \, B_n(Z + 2)
\end{equation}
and taking into account the identity of the electrons: the integrated phase space of the $0 \nu (2 \nu) \beta \beta$ decays contains a statistical factor of $1/2!$, which is not present in the case of the $0 \nu (2 \nu) \mathrm{EP} \beta$ decay modes since the bound and free electrons occupy complementary regions of the phase space. The corresponding rule for the integrated phase space reads:
\begin{align}
\label{eq:fb2}
& \frac{1}{2!} \int \frac{\mathrm{d} \mathbf{p}}{(2 \pi)^3} \, \frac{\mathrm{d} \mathbf{p}'}{(2 \pi)^3} \, F(Z + 2, \, E) \, F(Z + 2, \, E') \nonumber \\
& \longmapsto \frac{1}{4 \pi} \int \frac{\mathrm{d}^3 \mathbf{p}}{(2 \pi)^3} \, F(Z + 2, \, E) \sum_{n = n_{\mathrm{min}}}^{\infty} B_n(Z + 2).
\end{align}
In the bound-state double-$\beta$ decays, the binding energy of the produced electron $\lesssim 10 \, \mathrm{eV}$ can be neglected. Such an approximation does not affect the required accuracy but greatly simplifies the computation since the infinite sum of integrals in Eq.~(\ref{eq:fb2}) is factorized into the Fermi sum $\sum_{n = n_{\mathrm{min}}}^{\infty} B_n(Z + 2)$ and just one integral independent of $n$. The energy conservation in the $0 \nu \mathrm{EP} \beta$ decay implies that the free electron carries away the entire energy released in the decay: $E = m_e + Q$, whereas in the $2 \nu \mathrm{EP} \beta$ decay the energy is distributed between the electron and two antineutrinos: $\omega_2 = m_e + Q - E - \omega_1$.

\section{Bound-state wave functions of electrons in Dirac--Hartree--Fock method}
\label{sec:dhf}
The multiconfiguration Dirac--Hartree--Fock package \textsc{Grasp2K} solves the stationary $N$-body Dirac equation with the separable central atomic Hamiltonian \cite{Jon13}:
\begin{equation}
\sum_{i = 1}^N \left[ -i \boldsymbol{\nabla}_i \cdot \boldsymbol{\alpha} + m_e \, \beta - \frac{\alpha Z}{r_i} + V(r_i) \right] \Psi = E \Psi,
\end{equation}
where $\boldsymbol{\alpha} = \gamma^0 \, \boldsymbol{\gamma}$ and $\beta = \gamma^0$. The first two terms are followed by the potential-energy terms which account for the electron-nucleus Coulomb attraction and electron-electron Coulomb repulsion, respectively, where the latter is approximated by the mean field $V(r_i)$ generated by the surrounding electron cloud. The separability ensures that the energy eigenvalues are additive: $E = \sum_{i = 1}^N E_i$, while the many-electron wave functions are expressed in terms of the Slater determinants:
\begin{equation}
\Psi = \frac{1}{\sqrt{N!}} \sum_P (-1)^P \prod_{i = 1}^N \psi_{P(i)}(\mathbf{r}_i),
\end{equation}
where $\psi_i(\mathbf{r}_j) = \psi_{n_i \kappa_i \mu_i}(\mathbf{r}_j)$ and $P$ is the permutation of quantum numbers with parity $(-1)^P$. The nuclear part of the total wave function is disregarded by virtue of the Born--Oppenheimer approximation. The self-consistent field procedure then varies the radial functions $f_{n \kappa}(r)$ and $g_{n \kappa}(r)$ in iterative cycles until convergence is achieved.

The radial functions $f_{n, -1}(R)$ and $g_{n, +1}(R)$ are computed in the nuclear Coulomb potential of the daughter nucleus $\ce{_{Z + 2}^AY}$ for the ground-state electron configuration of the parent atom $\ce{_Z^AX}$ with an additional $\beta$-decay electron occupying an empty orbit. Since the convergence cannot be always guaranteed and the program only provides the electron-shell wave functions up to $n = 9$, we employ a combined approach:
\begin{enumerate}
\item The radial functions $f_{n, -1}(R)$ and $g_{n, +1}(R)$ are calculated based on initial estimates provided by the Thomas--Fermi model.
\item If the convergence cannot be achieved within a specified number of iterations, the radial functions $f_{n, -1}(R)$ and $g_{n, +1}(R)$ are calculated based on initial estimates provided by the non-relativistic Hartree--Fock approximation.
\item If both methods fail for the charge $Z$, we are looking for the values of $Z' \ne Z$ for which the calculation can be completed. The squares of the radial functions are then determined by fitting the available values for a fixed orbit using the power-law function: $f_{n, -1}^2(R), \, g_{n, +1}^2(R) \approx a Z^b$.
\item Finally, the squares of the radial functions with the principal quantum numbers above $n = 9$ are estimated for a given isotope from a fit of the available values for $n \le 9$ using the power-law function: $f_{n, -1}^2(R), \, g_{n, +1}^2(R) \approx c n^d$.
\end{enumerate}
In the atomic spectroscopy, power functions are often used to fit the dependence of observables on the atomic number $Z$ (see, e.g., \cite{Nis11}). On the other hand, the power law of the principal quantum number $n$ is motivated by the fact that, in the absence of shielding, the squares of nonrelativistic radial functions $n\mathrm{s}_{1/2}$ decrease at the origin as: $R_{n0}^2(0) \propto n^{-3}$. The simple power law enables us to explicitly perform the summation in $\sum_{n = n_{\mathrm{min}}}^{\infty} B_n(Z + 2)$ over the vacancies in the electron shell. The sum is expressed in terms of the Riemann zeta function $\zeta(z) = \sum_{n = 1}^{\infty} 1/n^z$. In average, the radial functions with $n > 9$ contribute to the decay rates at the level of only $\approx 4\%$ of the total value.

\begin{figure}[t]
\includegraphics[width=\columnwidth]{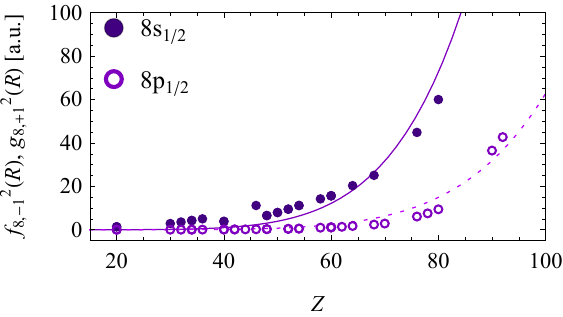}
\caption{\label{fig:g2kz}The squared radial wave functions $f_{n, -1}^2(R)$ and $g_{n, +1}^2(R)$ (in atomic units) for the subshells $8\mathrm{s}_{1/2}$ and $8\mathrm{p}_{1/2}$, respectively, fitted by the power function $a Z^b$ of the initial atomic number $Z$. The points represent the predictions of \textsc{Grasp2K}. The parameters determined from the fit read: $a = 1.1 \times 10^{-10}$, $b = 6.2$ ($8\mathrm{s}_{1/2}$) and $a = 8.1 \times 10^{-12}$, $b = 6.4$ ($8\mathrm{p}_{1/2}$).}
\end{figure}

\begin{figure}[t]
\includegraphics[width=\columnwidth]{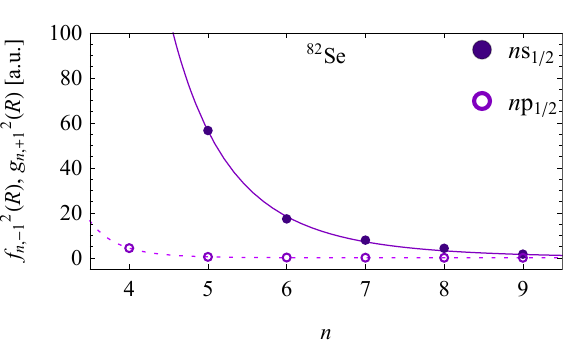}
\caption{\label{fig:g2kn}The squared radial wave functions $f_{n, -1}^2(R)$ and $g_{n, +1}^2(R)$ (in atomic units) for the isotope $\ce{^{82}Se}$ fitted by the power function $c n^d$ of the principal quantum number $n$. The parameters determined from the fit read: $c = 1.1 \times 10^6$, $d = -6.1$ ($n\mathrm{s}_{1/2}$) and $c = 4.6 \times 10^6$, $d = -10$ ($n\mathrm{p}_{1/2}$).}
\end{figure}

Figure~\ref{fig:g2kz} shows the results for a power-law fitting of the squared radial functions $f_{n, -1}^2(R)$ and $g_{n, +1}^2(R)$ at the nuclear radius $r = R$ as functions of the initial nuclear charge $Z$. The example of the $8\mathrm{s}_{1/2}$ and $8\mathrm{p}_{1/2}$ subshells is considered. The convergence cannot be achieved for all nuclei. The power-law dependence is in excellent agreement with the observed behavior of the computed radial wave functions. The radial functions at the nuclear radius $r = R$ as functions of the principal quantum number $n$ are shown in Fig.~\ref{fig:g2kn} for the isotope $\ce{^{82}Se}$. The results quoted in Figs.~\ref{fig:g2kz}--\ref{fig:g2kb} are presented in atomic units ($\mathrm{a.u.}$).

A simple qualitative explanation of the dependence of the bound-electron radial wave functions on $Z$ and $n$ at $r = R$ follows from the following considerations. The nodes of the radial part of a nonrelativistic wave function with $l = 0$ are localized partially outside the atom at $r \gtrsim 1$ (in $\mathrm{a.u.}$) and partially inside the atom at $r \lesssim 1$. The number of nodes inside the atom can be estimated for highly excited states using a semiclassical approximation, which is justified for $Z \gg 1$ and $r \lesssim 1$. At the boundary of the atom, the phase of the radial wave function is estimated to be: $\int_0^1 \mathrm{d}r \, \sqrt{2 [E - V(r)]} \sim Z^{1/3}$, so that the number of nodes inside the atom equals: $n_a \sim Z^{1/3}$. In the Coulomb potential, the squared radial wave function for small $r$ behaves like $\sim 1/n^3$. The atomic radius $\sim 1$ is small compared to the average radius $\sim n^2$ of the bound $\beta$-electron. The ratio $R_{n0}(1)/R_{n0}(0)$ is independent of $n$ for large $n$ and tends to $0.283$ at the infinity. Since $n_a$ nodes moved inside the atom, the square of the wave function at the atomic boundary becomes: $R_{n0}^2(1) \sim 1/(n - n_a)^3$. The matching at $r \sim 1$ of the outer part of the wave function with the semiclassical wave function at $r \lesssim 1$ leads to the appearance at $r \sim 0$ of an additional factor $Z$ (see, e.g., \cite{Lan77}), so finally:
\begin{equation}
\label{eq:q}
R_{n0}^2(0) \propto \frac{Z}{(n - n_a)^3}.
\end{equation}

The same result follows from the requirement of orthogonality of the wave function of the bound $\beta$-electron to the electron wave functions in the atom. The number of electrons occupying the atomic levels up to the principal quantum number $n_s$ with a completely filled outer shell is expressed as follows:
\begin{equation}
Z = \sum_{n = 1}^{n_s} \sum_{l = 0}^{n - 1} 2 \, (2l + 1) = \frac{1}{3} \, n_s \, (2 n_s + 1) (n_s + 1).
\end{equation}
In agreement with the semiclassical arguments given above, $n_s \sim (3Z/2)^{1/3}$. To ensure orthogonality, the bound $\beta$-electron should have one more node inside the atom compared to $n_s - 1$. One can verify that for $n_a \sim n_s$ Eq.~(\ref{eq:q}) reproduces the qualitative behavior of the upper radial function for $r = R$. The dependence on $Z$ for $n = 6$, shown in Fig.~\ref{fig:g2kz}, appears reasonable for $Z \gtrsim 20$. In the case of $\ce{_{34}^{82}Se}$, shown in Fig.~\ref{fig:g2kn}, the approximation (\ref{eq:q}) works reasonably well for $n \gtrsim 7$. We remark that Eq.~(\ref{eq:q}) is justified for $n \gg n_a$ and $Z \sim n_a^3 \gg 1$. The need for detailed calculations of the electron shell structure based on advanced programs of quantum chemistry like \textsc{Grasp2K} is quite obvious. Our calculations are made for isolated atoms, so the results are applicable directly to gaseous substances such as krypton or xenon. We expect that the presented calculations yield reasonable estimates also for solids.

\section{Results and discussion}
\label{sec:rd}
In Table~\ref{tab:br}, the double-$\beta$-decaying isotopes $\ce{_Z^AX}$ are listed together with (a) the $Q$ values obtained from the recent evaluation of atomic masses \cite{Wan17}, (b) the Fermi sums $\sum_{n = n_{\mathrm{min}}}^{\infty} B_n(Z + 2)$ (in atomic units) computed using the \textsc{Grasp2K} package, (c) the phase-space factors $G^{0 \nu (2 \nu) \mathrm{EP} \beta}$ and $G^{0 \nu (2 \nu) \beta \beta}$ associated with the ground-state $0^+ \longrightarrow 0^+$ nuclear transitions, and (d) the decay-rate ratios:
\begin{equation}
\frac{\Gamma^{0 \nu (2 \nu) \mathrm{EP} \beta}}{\Gamma^{0 \nu (2 \nu) \beta \beta}} \approx \frac{G^{0 \nu (2 \nu) \mathrm{EP} \beta}}{G^{0 \nu (2 \nu) \beta \beta}}
\end{equation}
which are independent of the NMEs and $m_{\beta \beta}$, and hence are free of uncertainties inherent in the nuclear-structure models and neutrino masses.

\begin{figure}[t]
\includegraphics[width=\columnwidth]{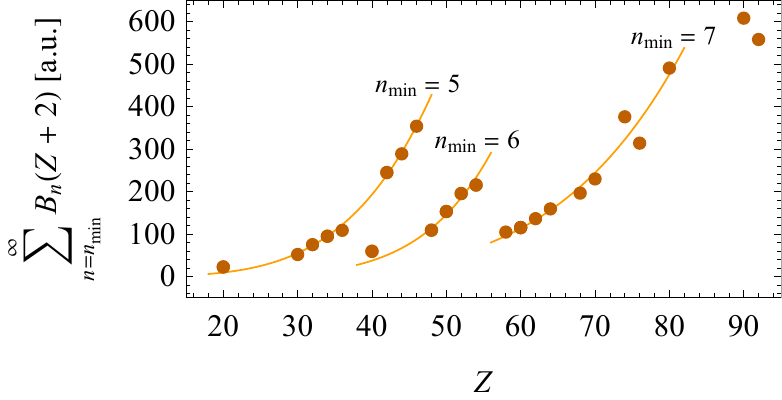}
\caption{\label{fig:g2kb}The Fermi sum $\sum_{n = n_{\mathrm{min}}}^{\infty} B_n(Z + 2)$ (in atomic units) as a function of the initial atomic number $Z$ of the parent nucleus for $n_{\mathrm{min}} = 5, \, 6, \, 7$.}
\end{figure}

\begin{figure}[t]
\includegraphics[width=\columnwidth]{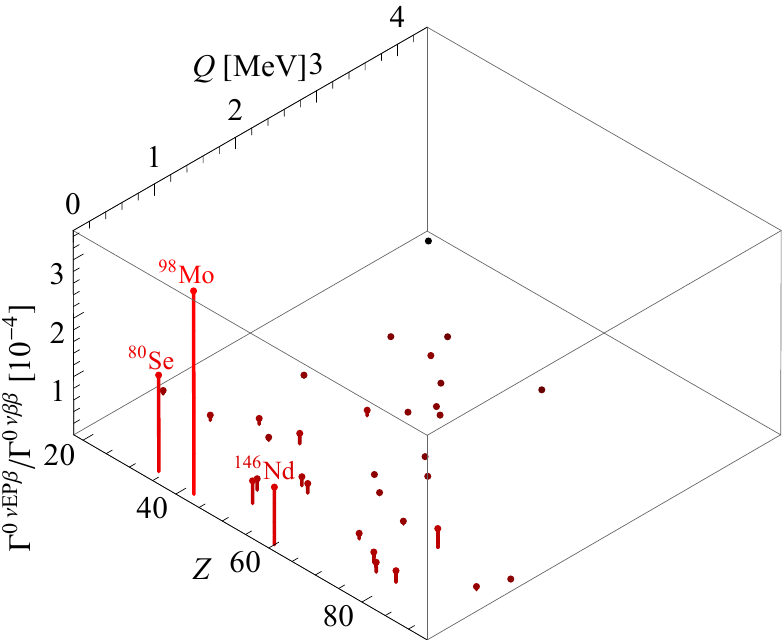}
\caption{\label{fig:br0n}The decay-rate ratio $\Gamma^{0 \nu \mathrm{EP} \beta}/\Gamma^{0 \nu \beta \beta}$ as a function of the atomic number $Z$ of the parent nucleus and the $Q$ value. The $0 \nu \mathrm{EP} \beta$ decay rate is maximal for the isotopes: $\ce{^{98}Mo}$, $\ce{^{80}Se}$ and $\ce{^{146}Nd}$, and decreases rapidly with increasing both $Z$ and $Q$.}
\end{figure}

\begin{figure}[t]
\includegraphics[width=\columnwidth]{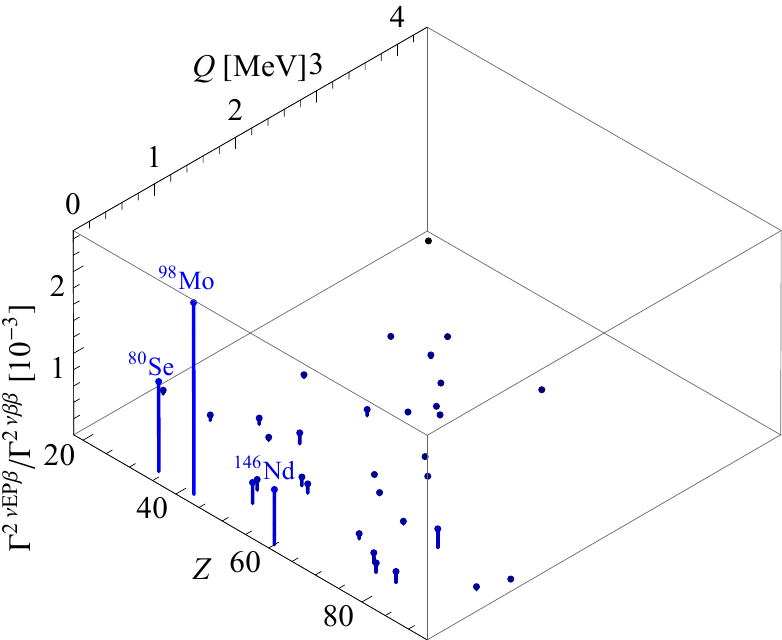}
\caption{\label{fig:br2n}The decay-rate ratio $\Gamma^{2 \nu \mathrm{EP} \beta}/\Gamma^{2 \nu \beta \beta}$ as a function of the atomic number $Z$ of the parent nucleus and the $Q$ value. The two-neutrino mode exhibits behavior similar to the neutrinoless mode (see Fig.~\ref{fig:br0n}), but the absolute values are by one order of magnitude higher.}
\end{figure}

The Fermi sum $\sum_{n = n_{\mathrm{min}}}^{\infty} B_n(Z + 2)$, shown in Fig.~\ref{fig:g2kb} (in atomic units), increases with the initial atomic number $Z$ and drops whenever the valence shell becomes fully occupied; the very large value of $2.199 \times 10^3$ for the isotope $\ce{_{78}^{198}Pt}$ with $n_{\mathrm{min}} = 6$ is out of bounds of the plot. The decay-rate ratios $\Gamma^{0 \nu (2 \nu) \mathrm{EP} \beta}/\Gamma^{0 \nu (2 \nu) \beta \beta}$, shown in Figs.~\ref{fig:br0n}--\ref{fig:br2n}, achieve their maximum for the isotopes with very low $Q$ values: $\ce{^{98}Mo}$, $\ce{^{80}Se}$ and $\ce{^{146}Nd}$, and decrease with increasing both $Z$ and $Q$. The two-neutrino channels exhibit decay-rate ratios by one order of magnitude higher than the neutrinoless channels. The overall suppression is mainly attributed to the presence of other electrons in the atom: the low-lying electron states (which would otherwise provide a dominant contribution) are already occupied, while the shielding effect of nuclear charge substantially reduces the bound-electron wave functions on the surface of the nucleus.

\begin{table*}
\caption{\label{tab:br}The double-$\beta$-decaying isotopes $\ce{_Z^AX}$ with the $Q$ values from Ref.~\cite{Wan17}, the Fermi sums $\sum_{n = n_{\mathrm{min}}}^{\infty} B_n(Z + 2)$ (in atomic units) over the vacant electron shells of the daughter ion, the bound-state decay phase-space factors $G^{0 \nu (2 \nu) \mathrm{EP} \beta}$ from Eqs.~(\ref{eq:g0n})--(\ref{eq:g2n}), the standard phase-space factors $G^{0 \nu (2 \nu) \beta \beta}$, and the relative frequencies of bound-state to continuum-state decays $\Gamma^{0 \nu (2 \nu) \mathrm{EP} \beta}/\Gamma^{0 \nu (2 \nu) \beta \beta}$.}
\begin{ruledtabular}
\begin{tabular}{cd{1.3}d{1.8}d{1.10}d{1.10}d{1.8}d{1.10}d{1.10}d{1.8}}
$\ce{_Z^AX}$ & \multicolumn{1}{c}{$Q \, [\mathrm{MeV}]$} & \multicolumn{1}{c}{$\sum_n B_n \, [\mathrm{a.u.}]$} & \multicolumn{1}{c}{$G^{0 \nu \mathrm{EP} \beta} \, [\mathrm{yr}^{-1}]$} & \multicolumn{1}{c}{$G^{0 \nu \beta \beta} \, [\mathrm{yr}^{-1}]$} & \multicolumn{1}{c}{$\Gamma^{0 \nu \mathrm{EP} \beta}/\Gamma^{0 \nu \beta \beta}$} & \multicolumn{1}{c}{$G^{2 \nu \mathrm{EP} \beta} \, [\mathrm{yr}^{-1}]$} & \multicolumn{1}{c}{$G^{2 \nu \beta \beta} \, [\mathrm{yr}^{-1}]$} & \multicolumn{1}{c}{$\Gamma^{2 \nu \mathrm{EP} \beta}/\Gamma^{2 \nu \beta \beta}$} \\
\hline
$\ce{_{20}^{46}Ca}$ & 0.988 & 2.246 \times 10^1 & 9.343 \times 10^{-22} & 1.499 \times 10^{-16} & 6.23 \times 10^{-6} & 2.262 \times 10^{-27} & 4.734 \times 10^{-23} & 4.78 \times 10^{-5} \\
$\ce{_{20}^{48}Ca}$ & 4.268 & 2.245 \times 10^1 & 9.227 \times 10^{-21} & 2.632 \times 10^{-14} & 3.51 \times 10^{-7} & 5.923 \times 10^{-23} & 1.594 \times 10^{-17} & 3.72 \times 10^{-6} \\
$\ce{_{30}^{70}Zn}$ & 0.997 & 5.180 \times 10^1 & 2.302 \times 10^{-21} & 2.463 \times 10^{-16} & 9.34 \times 10^{-6} & 8.521 \times 10^{-27} & 1.239 \times 10^{-22} & 6.88 \times 10^{-5} \\
$\ce{_{32}^{76}Ge}$ & 2.039 & 7.495 \times 10^1 & 9.491 \times 10^{-21} & 2.615 \times 10^{-15} & 3.63 \times 10^{-6} & 1.621 \times 10^{-24} & 5.280 \times 10^{-20} & 3.07 \times 10^{-5} \\
$\ce{_{34}^{80}Se}$ & 0.134 & 9.482 \times 10^1 & 7.822 \times 10^{-22} & 4.724 \times 10^{-18} & 1.66 \times 10^{-4} & 6.761 \times 10^{-32} & 6.119 \times 10^{-29} & 1.10 \times 10^{-3} \\
$\ce{_{34}^{82}Se}$ & 2.998 & 9.476 \times 10^1 & 2.263 \times 10^{-20} & 1.152 \times 10^{-14} & 1.97 \times 10^{-6} & 3.250 \times 10^{-23} & 1.779 \times 10^{-18} & 1.83 \times 10^{-5} \\
$\ce{_{36}^{86}Kr}$ & 1.257 & 1.087 \times 10^2 & 7.120 \times 10^{-21} & 6.798 \times 10^{-16} & 1.05 \times 10^{-5} & 1.068 \times 10^{-25} & 1.354 \times 10^{-21} & 7.88 \times 10^{-5} \\
$\ce{_{40}^{94}Zr}$ & 1.145 & 5.933 \times 10^1 & 3.736 \times 10^{-21} & 6.725 \times 10^{-16} & 5.56 \times 10^{-6} & 3.773 \times 10^{-26} & 9.254 \times 10^{-22} & 4.08 \times 10^{-5} \\
$\ce{_{40}^{96}Zr}$ & 3.356 & 5.928 \times 10^1 & 1.867 \times 10^{-20} & 2.440 \times 10^{-14} & 7.65 \times 10^{-7} & 5.714 \times 10^{-23} & 7.899 \times 10^{-18} & 7.23 \times 10^{-6} \\
$\ce{_{42}^{98}Mo}$ & 0.109 & 2.447 \times 10^2 & 2.358 \times 10^{-21} & 6.769 \times 10^{-18} & 3.48 \times 10^{-4} & 7.509 \times 10^{-32} & 3.198 \times 10^{-29} & 2.35 \times 10^{-3} \\
$\ce{_{42}^{100}Mo}$ & 3.034 & 2.445 \times 10^2 & 6.792 \times 10^{-20} & 1.890 \times 10^{-14} & 3.59 \times 10^{-6} & 1.255 \times 10^{-22} & 3.816 \times 10^{-18} & 3.29 \times 10^{-5} \\
$\ce{_{44}^{104}Ru}$ & 1.299 & 2.887 \times 10^2 & 2.343 \times 10^{-20} & 1.270 \times 10^{-15} & 1.84 \times 10^{-5} & 5.050 \times 10^{-25} & 3.676 \times 10^{-21} & 1.37 \times 10^{-4} \\
$\ce{_{46}^{110}Pd}$ & 2.017 & 3.537 \times 10^2 & 5.601 \times 10^{-20} & 5.778 \times 10^{-15} & 9.69 \times 10^{-6} & 1.284 \times 10^{-23} & 1.624 \times 10^{-19} & 7.91 \times 10^{-5} \\
$\ce{_{48}^{114}Cd}$ & 0.545 & 1.091 \times 10^2 & 3.520 \times 10^{-21} & 1.795 \times 10^{-16} & 1.96 \times 10^{-5} & 8.819 \times 10^{-28} & 6.703 \times 10^{-24} & 1.32 \times 10^{-4} \\
$\ce{_{48}^{116}Cd}$ & 2.813 & 1.089 \times 10^2 & 2.987 \times 10^{-20} & 2.064 \times 10^{-14} & 1.45 \times 10^{-6} & 4.243 \times 10^{-23} & 3.311 \times 10^{-18} & 1.28 \times 10^{-5} \\
$\ce{_{50}^{122}Sn}$ & 0.373 & 1.531 \times 10^2 & 3.682 \times 10^{-21} & 9.414 \times 10^{-17} & 3.91 \times 10^{-5} & 1.293 \times 10^{-28} & 4.986 \times 10^{-25} & 2.59 \times 10^{-4} \\
$\ce{_{50}^{124}Sn}$ & 2.291 & 1.527 \times 10^2 & 3.131 \times 10^{-20} & 1.132 \times 10^{-14} & 2.77 \times 10^{-6} & 1.577 \times 10^{-23} & 6.822 \times 10^{-19} & 2.31 \times 10^{-5} \\
$\ce{_{52}^{128}Te}$ & 0.867 & 1.953 \times 10^2 & 1.139 \times 10^{-20} & 7.291 \times 10^{-16} & 1.56 \times 10^{-5} & 3.634 \times 10^{-26} & 3.349 \times 10^{-22} & 1.09 \times 10^{-4} \\
$\ce{_{52}^{130}Te}$ & 2.528 & 1.952 \times 10^2 & 4.845 \times 10^{-20} & 1.810 \times 10^{-14} & 2.68 \times 10^{-6} & 4.327 \times 10^{-23} & 1.893 \times 10^{-18} & 2.29 \times 10^{-5} \\
$\ce{_{54}^{134}Xe}$ & 0.824 & 2.154 \times 10^2 & 1.251 \times 10^{-20} & 7.487 \times 10^{-16} & 1.67 \times 10^{-5} & 3.201 \times 10^{-26} & 2.776 \times 10^{-22} & 1.15 \times 10^{-4} \\
$\ce{_{54}^{136}Xe}$ & 2.458 & 2.152 \times 10^2 & 5.349 \times 10^{-20} & 1.883 \times 10^{-14} & 2.84 \times 10^{-6} & 4.310 \times 10^{-23} & 1.795 \times 10^{-18} & 2.40 \times 10^{-5} \\
$\ce{_{58}^{142}Ce}$ & 1.417 & 1.046 \times 10^2 & 1.353 \times 10^{-20} & 4.564 \times 10^{-15} & 2.96 \times 10^{-6} & 6.332 \times 10^{-25} & 2.873 \times 10^{-20} & 2.20 \times 10^{-5} \\
$\ce{_{60}^{146}Nd}$ & 0.070 & 1.152 \times 10^2 & 1.886 \times 10^{-21} & 1.907 \times 10^{-17} & 9.89 \times 10^{-5} & 6.262 \times 10^{-33} & 9.236 \times 10^{-30} & 6.78 \times 10^{-4} \\
$\ce{_{60}^{148}Nd}$ & 1.928 & 1.151 \times 10^2 & 2.398 \times 10^{-20} & 1.358 \times 10^{-14} & 1.77 \times 10^{-6} & 5.933 \times 10^{-24} & 4.253 \times 10^{-19} & 1.40 \times 10^{-5} \\
$\ce{_{60}^{150}Nd}$ & 3.371 & 1.150 \times 10^2 & 5.437 \times 10^{-20} & 8.829 \times 10^{-14} & 6.16 \times 10^{-7} & 2.700 \times 10^{-22} & 4.815 \times 10^{-17} & 5.61 \times 10^{-6} \\
$\ce{_{62}^{154}Sm}$ & 1.251 & 1.361 \times 10^2 & 1.685 \times 10^{-20} & 4.413 \times 10^{-15} & 3.82 \times 10^{-6} & 4.478 \times 10^{-25} & 1.617 \times 10^{-20} & 2.77 \times 10^{-5} \\
$\ce{_{64}^{160}Gd}$ & 1.731 & 1.592 \times 10^2 & 3.198 \times 10^{-20} & 1.336 \times 10^{-14} & 2.39 \times 10^{-6} & 4.892 \times 10^{-24} & 2.658 \times 10^{-19} & 1.84 \times 10^{-5} \\
$\ce{_{68}^{170}Er}$ & 0.655 & 1.963 \times 10^2 & 1.464 \times 10^{-20} & 1.513 \times 10^{-15} & 9.68 \times 10^{-6} & 1.442 \times 10^{-26} & 2.202 \times 10^{-22} & 6.55 \times 10^{-5} \\
$\ce{_{70}^{176}Yb}$ & 1.085 & 2.297 \times 10^2 & 3.150 \times 10^{-20} & 6.129 \times 10^{-15} & 5.14 \times 10^{-6} & 4.633 \times 10^{-25} & 1.272 \times 10^{-20} & 3.64 \times 10^{-5} \\
$\ce{_{74}^{186}W}$ & 0.491 & 3.759 \times 10^2 & 2.789 \times 10^{-20} & 1.508 \times 10^{-15} & 1.85 \times 10^{-5} & 6.473 \times 10^{-27} & 5.220 \times 10^{-23} & 1.24 \times 10^{-4} \\
$\ce{_{76}^{192}Os}$ & 0.406 & 3.139 \times 10^2 & 2.200 \times 10^{-20} & 1.292 \times 10^{-15} & 1.70 \times 10^{-5} & 1.881 \times 10^{-27} & 1.651 \times 10^{-23} & 1.14 \times 10^{-4} \\
$\ce{_{78}^{198}Pt}$ & 1.050 & 2.199 \times 10^3 & 3.976 \times 10^{-19} & 1.231 \times 10^{-14} & 3.23 \times 10^{-5} & 5.701 \times 10^{-24} & 2.503 \times 10^{-20} & 2.28 \times 10^{-4} \\
$\ce{_{80}^{204}Hg}$ & 0.420 & 4.906 \times 10^2 & 4.237 \times 10^{-20} & 2.121 \times 10^{-15} & 2.00 \times 10^{-5} & 4.630 \times 10^{-27} & 3.456 \times 10^{-23} & 1.34 \times 10^{-4} \\
$\ce{_{90}^{232}Th}$ & 0.837 & 6.081 \times 10^2 & 1.508 \times 10^{-19} & 2.696 \times 10^{-14} & 5.59 \times 10^{-6} & 8.012 \times 10^{-25} & 2.070 \times 10^{-20} & 3.87 \times 10^{-5} \\
$\ce{_{92}^{238}U}$ & 1.145 & 5.579 \times 10^2 & 2.058 \times 10^{-19} & 6.981 \times 10^{-14} & 2.95 \times 10^{-6} & 6.096 \times 10^{-24} & 2.902 \times 10^{-19} & 2.10 \times 10^{-5} \\
\end{tabular}
\end{ruledtabular}
\end{table*}

\begin{figure}[t]
\includegraphics[width=\columnwidth]{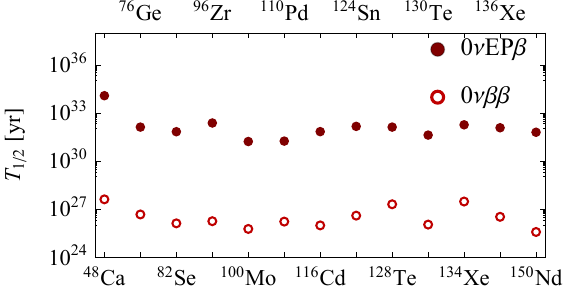}
\caption{\label{fig:hl0n}The half-lives $T_{1/2}^{0 \nu \mathrm{EP} \beta}$ and $T_{1/2}^{0 \nu \beta \beta}$ for the isotopes with the calculated NMEs \cite{Sim13,Fan15} estimated assuming the unquenched axial coupling constant: $g_A = 1.27$ and the effective Majorana neutrino mass: $\left| m_{\beta \beta} \right| = 50 \, \mathrm{meV}$.}
\end{figure}

\begin{figure}[t]
\includegraphics[width=\columnwidth]{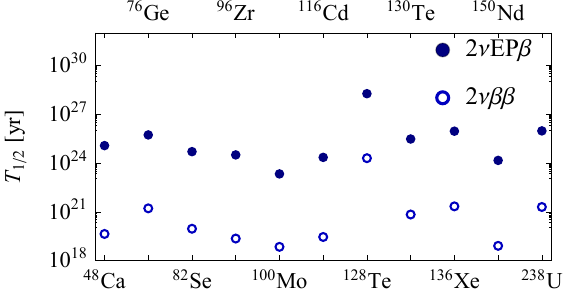}
\caption{\label{fig:hl2n}The half-lives $T_{1/2}^{2 \nu \mathrm{EP} \beta}$ and $T_{1/2}^{2 \nu \beta \beta}$ \cite{Bar15} for the isotopes with double-$\beta$ decays observed experimentally. The $2 \nu \mathrm{EP} \beta$ and $0 \nu \beta \beta$ decay rates are comparable in magnitude.}
\end{figure}

In Table~\ref{tab:hl}, the double-$\beta$-decaying isotopes $\ce{_Z^AX}$ with available NMEs are listed together with their half-lives $T_{1/2}^{0 \nu (2 \nu) \mathrm{EP} \beta}$ and $T_{1/2}^{0 \nu (2 \nu) \beta \beta}$. The NMEs $M^{0 \nu \beta \beta}$ used for the estimates were obtained within the spherical pn-QRPA approach including the CD-Bonn nucleon-nucleon potential with short-range correlations and the partial isospin-symmetry restoration \cite{Sim13}, except for the isotope $\ce{^{150}Nd}$ which was treated separately within the deformed pn-QRPA model \cite{Fan15}. We estimate the neutrinoless half-lives assuming the unquenched value of the axial-vector coupling constant $g_A = 1.27$ and the effective Majorana neutrino mass at the top of the allowed inverted-hierarchy region: $\left| m_{\beta \beta} \right| = 50 \, \mathrm{meV}$. The half-lives $T_{1/2}^{2 \nu \mathrm{EP} \beta}$ are derived based on the values of $T_{1/2}^{2 \nu \beta \beta}$ measured experimentally \cite{Bar15}; these are further used to extract the NMEs listed for $g_A = 1.27$. While the $0 \nu \mathrm{EP} \beta$ decay mode is strongly suppressed and can hardly be experimentally observed in the near future, the half-lives of its $2 \nu \mathrm{EP} \beta$ counterpart are already comparable to the present sensitivity to the $0 \nu \beta \beta$ decay. Figures~\ref{fig:hl0n} and \ref{fig:hl2n} show the neutrinoless and two-neutrino double-$\beta$-decay half-lives for the isotopes listed in Table~\ref{tab:hl}.

\begin{table*}
\caption{\label{tab:hl}The double-$\beta$-decaying isotopes $\ce{_Z^AX}$ for which the NMEs were determined theoretically or experimentally \cite{Sim13,Fan15}, their corresponding half-lives $T_{1/2}^{0 \nu \mathrm{EP} \beta}$ and $T_{1/2}^{0 \nu \beta \beta}$ estimated for $g_A = 1.27$ and $\left| m_{\beta \beta} \right| = 50 \, \mathrm{meV}$, and $T_{1/2}^{2 \nu \mathrm{EP} \beta}$ derived from the measured values of $T_{1/2}^{2 \nu \beta \beta}$ \cite{Bar15}.}
\begin{ruledtabular}
\begin{tabular}{cd{1.3}d{1.8}d{1.8}d{1.9}d{1.8}d{1.8}}
$\ce{_Z^AX}$ & \multicolumn{1}{c}{$\left| M^{0 \nu \beta \beta} \right|$} & \multicolumn{1}{c}{$T_{1/2}^{0 \nu \mathrm{EP} \beta} \, [\mathrm{yr}]$} & \multicolumn{1}{c}{$T_{1/2}^{0 \nu \beta \beta} \, [\mathrm{yr}]$} & \multicolumn{1}{c}{$\left| m_e \, M^{2 \nu \beta \beta} \right|$} & \multicolumn{1}{c}{$T_{1/2}^{2 \nu \mathrm{EP} \beta} \, [\mathrm{yr}]$} & \multicolumn{1}{c}{$T_{1/2}^{2 \nu \beta \beta} \, [\mathrm{yr}]$} \\
\hline
$\ce{_{20}^{48}Ca}$ & 0.594 & 1.23 \times 10^{34} & 4.32 \times 10^{27} & 2.341 \times 10^{-2} & 1.18 \times 10^{25} & 4.40 \times 10^{19} \\
$\ce{_{32}^{76}Ge}$ & 5.571 & 1.36 \times 10^{32} & 4.95 \times 10^{26} & 6.642 \times 10^{-2} & 5.38 \times 10^{25} & 1.65 \times 10^{21} \\
$\ce{_{34}^{82}Se}$ & 5.018 & 7.05 \times 10^{31} & 1.38 \times 10^{26} & 4.846 \times 10^{-2} & 5.04 \times 10^{24} & 9.20 \times 10^{19} \\
$\ce{_{40}^{96}Zr}$ & 2.957 & 2.46 \times 10^{32} & 1.88 \times 10^{26} & 4.600 \times 10^{-2} & 3.18 \times 10^{24} & 2.30 \times 10^{19} \\
$\ce{_{42}^{100}Mo}$ & 5.850 & 1.73 \times 10^{31} & 6.21 \times 10^{25} & 1.191 \times 10^{-1} & 2.16 \times 10^{23} & 7.10 \times 10^{18} \\
$\ce{_{46}^{110}Pd}$ & 6.255 & 1.83 \times 10^{31} & 1.78 \times 10^{26} & & & \\
$\ce{_{48}^{116}Cd}$ & 4.343 & 7.13 \times 10^{31} & 1.03 \times 10^{26} & 6.360 \times 10^{-2} & 2.24 \times 10^{24} & 2.87 \times 10^{19} \\
$\ce{_{50}^{124}Sn}$ & 2.913 & 1.51 \times 10^{32} & 4.18 \times 10^{26} & & & \\
$\ce{_{52}^{128}Te}$ & 5.084 & 1.36 \times 10^{32} & 2.13 \times 10^{27} & 2.396 \times 10^{-2} & 1.84 \times 10^{28} & 2.00 \times 10^{24} \\
$\ce{_{52}^{130}Te}$ & 4.373 & 4.33 \times 10^{31} & 1.16 \times 10^{26} & 1.716 \times 10^{-2} & 3.02 \times 10^{25} & 6.90 \times 10^{20} \\
$\ce{_{54}^{134}Xe}$ & 4.119 & 1.89 \times 10^{32} & 3.16 \times 10^{27} & & & \\
$\ce{_{54}^{136}Xe}$ & 2.460 & 1.24 \times 10^{32} & 3.52 \times 10^{26} & 9.888 \times 10^{-3} & 9.12 \times 10^{25} & 2.19 \times 10^{21} \\
$\ce{_{60}^{150}Nd}$ & 3.367 & 6.51 \times 10^{31} & 4.01 \times 10^{25} & 3.120 \times 10^{-2} & 1.46 \times 10^{24} & 8.20 \times 10^{18} \\
$\ce{_{92}^{238}U}$ & & & & 2.573 \times 10^{-2} & 9.52 \times 10^{25} & 2.00 \times 10^{21} \\
\end{tabular}
\end{ruledtabular}
\end{table*}

\begin{figure}[t]
\includegraphics[width=\columnwidth]{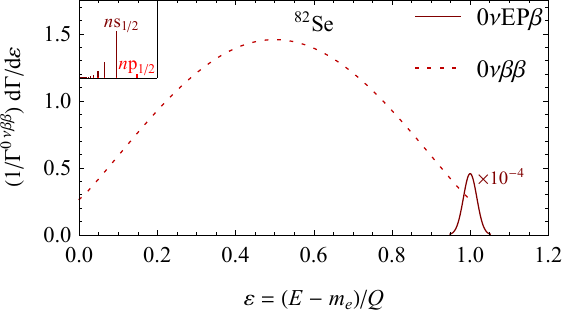}
\caption{\label{fig:ses0n}The $0 \nu \mathrm{EP} \beta$ and $0 \nu \beta \beta$ one-electron spectra $(1/\Gamma^{0 \nu \beta \beta}) \, \mathrm{d}\Gamma/\mathrm{d}\varepsilon$ as functions of the normalized electron kinetic energy $\varepsilon = (E - m_e)/Q$ for the isotope $\ce{^{82}Se}$. The $0 \nu \mathrm{EP} \beta$ peak is represented by a Gaussian with $\mathrm{FWHM}/Q = 7\%/\sqrt{Q/\mathrm{MeV}}$, which corresponds to the planned energy resolution of the SuperNEMO calorimeters, and scaled by a factor of $10^4$. The composition of the $0 \nu \mathrm{EP} \beta$ peak beyond the endpoint $\varepsilon = 1$ is shown in the upper left corner.}
\end{figure}

\begin{figure}[t]
\includegraphics[width=\columnwidth]{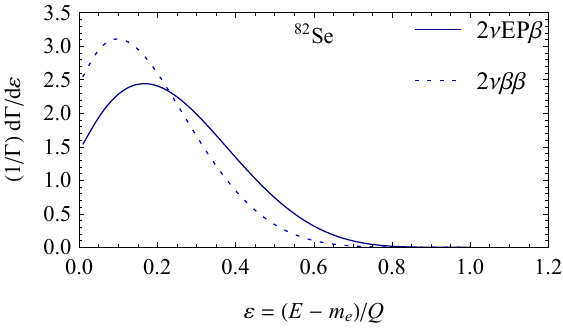}
\caption{\label{fig:ses2n}The $2 \nu \mathrm{EP} \beta$ and $2 \nu \beta \beta$ one-electron spectra $(1/\Gamma) \, \mathrm{d}\Gamma/\mathrm{d}\varepsilon$ as functions of the normalized electron kinetic energy $\varepsilon = (E - m_e)/Q$ for $\ce{^{82}Se}$.}
\end{figure}

The $0 \nu (2 \nu) \mathrm{EP} \beta$ and $0 \nu (2 \nu) \beta \beta$ one-electron spectra are described by the differential decay rates $(1/\Gamma) \, \mathrm{d}\Gamma/\mathrm{d}\varepsilon$, conventionally normalized to unity and expressed as functions of the dimensionless portion of the electron kinetic energy $\varepsilon = (E - m_e)/Q$:
\begin{align}
& \frac{\mathrm{d}\Gamma^{0 \nu \mathrm{EP} \beta}}{\mathrm{d}\varepsilon} = g_A^4 \, \frac{G_{\beta}^4 \, m_e^2}{32 \pi^4 R^2} \left| M^{0 \nu \beta \beta} \right|^2 \left| \frac{m_{\beta \beta}}{m_e} \right|^2 Q \nonumber \\
& \times \sum_{n = n_{\mathrm{min}}}^{\infty} B_n(Z + 2) \, F(Z + 2, \, E) \, E \, p \, \delta(m_e + Q - E),
\end{align}
\begin{align}
& \frac{\mathrm{d}\Gamma^{2 \nu \mathrm{EP} \beta}}{\mathrm{d}\varepsilon} = g_A^4 \, \frac{G_{\beta}^4}{8 \pi^6 m_e^2} \left| m_e \, M^{2 \nu \beta \beta} \right|^2 Q \nonumber \\
& \times \sum_{n = n_{\mathrm{min}}}^{\infty} B_n(Z + 2) \, F(Z + 2, \, E) \, E \, p \int\limits_0^{(1 - \varepsilon) Q} \mathrm{d}\omega_1 \, \omega_1^2 \, \omega_2^2.
\end{align}
Shown in Figs.~\ref{fig:ses0n}--\ref{fig:ses2n} are the one-electron spectra for the neutrinoless and two-neutrino double-$\beta$ decays of $\ce{_{34}^{82}Se}$. The $0 \nu \mathrm{EP} \beta$ peak consists of a large number of discrete contributions, each shifted above the $Q$ value by the electron binding energy ($\lesssim 10 \, \mathrm{eV}$); however, these are indistinguishable under any realistic energy resolution. The $2 \nu \mathrm{EP} \beta$ spectrum covers the entire energy range, which could lead to a slight deformation of the measured $2 \nu \beta \beta$ data.

The one-electron spectra are studied with unprecedented accuracy in the tracking-and-calorimetry double-$\beta$ decay experiments based on the external-source technique at the Modane Underground Laboratory (LSM). The NEMO-3 detector \cite{Arn15}, which operated during 2003--2011, exploited a cylindrical geometry and observed more than $7 \times 10^5$ positive $2 \nu \beta \beta$ events with a high signal-to-background ratio for $7 \, \mathrm{kg}$ of its primary source isotope $\ce{^{100}Mo}$ during $3.5 \, \mathrm{yr}$ of data taking (the low-radon phase) \cite{Bar13}. The next-generation detector SuperNEMO \cite{Arn10}, which is currently under construction, will deploy the source modules comprising 20 thin foils totalling in $100 \, \mathrm{kg}$ of enriched and purified $\ce{^{82}Se}$, with possible addition of $\ce{^{48}Ca}$ or $\ce{^{150}Nd}$ isotopes. The tracking chamber will consist of nine planar high-granularity drift cells operating in Geiger regime in a magnetic field of $2.5 \, \mathrm{mT}$, and thus enable charge-sign particle identification and vertex reconstruction, secure enhanced background rejection, and provide means to study angular correlations in addition to the one-electron spectra. The calorimeter walls will be composed of segmented low-$Z$ organic-scintillator blocks connected to photomultiplier tubes, striving to achieve the energy resolution: $\mathrm{FWHM}/Q = 7\%/\sqrt{Q/\mathrm{MeV}}$ in the region of interest (ROI) $2.8 \textrm{--} 3.2 \, \mathrm{MeV}$ around the endpoint $Q \approx 2.998 \, \mathrm{MeV}$. The first planar SuperNEMO module ``Demonstrator'' with $7 \, \mathrm{kg}$ of the source isotope $\ce{^{82}Se}$ is currently in its final stages of the development.

\begin{figure}[t]
\includegraphics[width=\columnwidth]{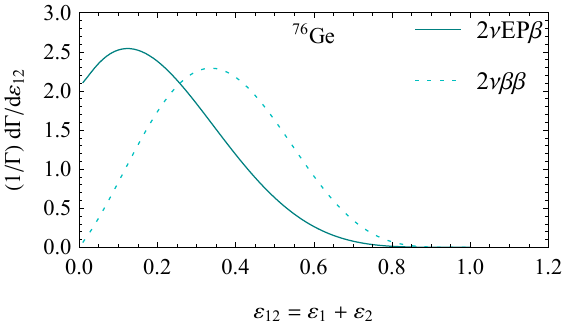}
\caption{\label{fig:tes2n}The $2 \nu \mathrm{EP} \beta$ and $2 \nu \beta \beta$ normalized differential decay rates $(1/\Gamma) \, \mathrm{d}\Gamma/\mathrm{d}\varepsilon_{12}$ as functions of the sum $\varepsilon_{12} = \varepsilon_1 + \varepsilon_2$ of electron energies for $\ce{^{76}Ge}$.}
\end{figure}

While the calorimetric measurements are unable to distinguish between the $0 \nu \mathrm{EP} \beta$ and $0 \nu \beta \beta$ peaks, the $2 \nu \mathrm{EP} \beta$ decay mode can also be identified by studying the two-electron spectra, which measure the total energy deposited by the emitted electrons. The normalized $2 \nu \beta \beta$ differential decay rate $(1/\Gamma^{2 \nu \beta \beta}) \, \mathrm{d}\Gamma^{2 \nu \beta \beta}/\mathrm{d}\varepsilon_{12}$ expressed as a function of the sum of electron kinetic energies $\varepsilon_1 = (E_1 - m_e)/Q$ and $\varepsilon_2 = (E_2 - m_e)/Q$ can be derived from the standard $2 \nu \beta \beta$ one-electron energy distribution via the substitutions $\varepsilon_{12} = \varepsilon_1 + \varepsilon_2$ and $\rho = \varepsilon_1/(\varepsilon_1 + \varepsilon_2)$:
\begin{align}
\frac{\mathrm{d}\Gamma^{2 \nu \beta \beta}}{\mathrm{d}\varepsilon_{12}} & = g_A^4 \frac{G_{\beta}^4}{8 \pi^7 m_e^2} \left| m_e \, M^{2 \nu \beta \beta} \right|^2 Q^2 \varepsilon_{12} \nonumber \\
& \quad \times \int\limits_0^1 \mathrm{d}\rho \, F(Z + 2, \, E_1) \, E_1 \, p_1 \, F(Z + 2, \, E_2) \, E_2 \, p_2 \nonumber \\
& \quad \times \int\limits_0^{(1 - \varepsilon_{12}) Q} \mathrm{d}\omega_1 \, \omega_1^2 \, \omega_2^2,
\end{align}
where $E_1$, $E_2$ and $p_1$, $p_2$ are the energies and momenta of the $\beta$-electrons and the energy conservation yields: $\omega_2 = (1 - \varepsilon_{12}) Q - \omega_1$. The spectral shapes of the $2 \nu \mathrm{EP} \beta$ and $2 \nu \beta \beta$ decays are shown in Fig.~\ref{fig:tes2n}. Since the two-electron spectra are usually measured with much higher event rates and less complicated background, a significant $2 \nu \mathrm{EP} \beta$ discovery potential is expected in the calorimetric double-$\beta$-decay experiments, in particular, CUORE ($\ce{^{130}Te}$) \cite{Ald18}, EXO-200 ($\ce{^{136}Xe}$) \cite{Alb18} and GERDA ($\ce{^{76}Ge}$) \cite{Ago18}.

\begin{figure}[t]
\includegraphics[width=\columnwidth]{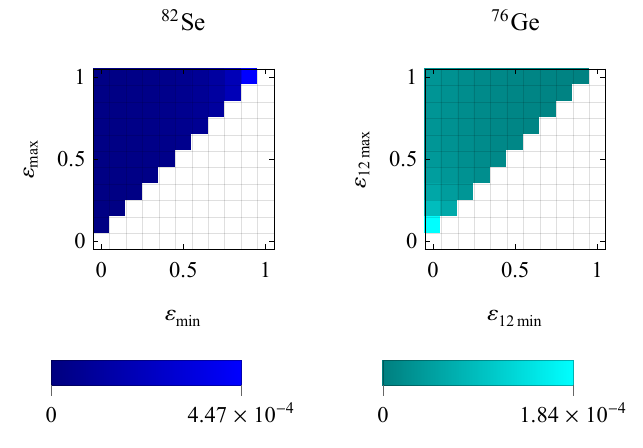}
\caption{\label{fig:roi2n}The ratios (\ref{eq:roi}) between the integrated $2 \nu \mathrm{EP} \beta$ and $2 \nu \beta \beta$ decay rates as functions of the energy intervals $[\varepsilon_{\mathrm{min}}, \, \varepsilon_{\mathrm{max}}]$ and $[\varepsilon_{12\mathrm{min}}, \, \varepsilon_{12\mathrm{max}}]$. The one-electron ratio is given for $\ce{^{82}Se}$ (left panel) and the two-electron ratio refers to $\ce{^{76}Ge}$ (right panel). The ROIs with the highest $2 \nu \mathrm{EP} \beta$ sensitivity belong to the opposite sides of the $\varepsilon_{(12)}$-interval at $\varepsilon = 1$ and $\varepsilon_{12} = 0$, respectively.}
\end{figure}

For data analysis, it is often desirable to specify the ratios:
\begin{align}
\label{eq:roi}
& \frac{\int_{\varepsilon_{\mathrm{min}}}^{\varepsilon_{\mathrm{max}}} \mathrm{d}\varepsilon \, (\mathrm{d}\Gamma^{2 \nu \mathrm{EP} \beta}/\mathrm{d}\varepsilon)}{\int_{\varepsilon_{\mathrm{min}}}^{\varepsilon_{\mathrm{max}}} \mathrm{d}\varepsilon \, (\mathrm{d}\Gamma^{2 \nu \beta \beta}/\mathrm{d}\varepsilon)}, \nonumber \\
& \frac{\int_{\varepsilon_{12\mathrm{min}}}^{\varepsilon_{12\mathrm{max}}} \mathrm{d}\varepsilon_{12} \, (\mathrm{d}\Gamma^{2 \nu \mathrm{EP} \beta}/\mathrm{d}\varepsilon_{12})}{\int_{\varepsilon_{12\mathrm{min}}}^{\varepsilon_{12\mathrm{max}}} \mathrm{d}\varepsilon_{12} \, (\mathrm{d}\Gamma^{2 \nu \beta \beta}/\mathrm{d}\varepsilon_{12})}
\end{align}
between the integrated $2 \nu \mathrm{EP} \beta$ and $2 \nu \beta \beta$ decay rates as functions of the energy intervals $[\varepsilon_{\mathrm{min}}, \, \varepsilon_{\mathrm{max}}]$ and $[\varepsilon_{12\mathrm{min}}, \, \varepsilon_{12\mathrm{max}}]$ in order to identify the ROIs in which the $2 \nu \mathrm{EP} \beta$ decay is best visible relative to its $2 \nu \beta \beta$ counterpart. While the one-electron ratios are maximal in a small ROI at the spectrum endpoint $Q$, the two-electron ratios reveal the highest $2 \nu \mathrm{EP} \beta$ sensitivity near the opposite end of the energy domain. In these ROIs, the $2 \nu \mathrm{EP} \beta$ decay mode could for the given isotopes account for as much as $\sim 100 \, \mathrm{ppm}$ of the registered events. The ratios from Eq.~(\ref{eq:roi}) for the one- and two-electron spectra associated with the decays of $\ce{^{82}Se}$ and $\ce{^{76}Ge}$, respectively, are shown in Fig.~\ref{fig:roi2n}.

At temperatures $T \gg \alpha^2 Z^2 m_e \sim 10^8 (Z/34)^2 \, \mathrm{K}$, atoms become fully ionized and the $\beta$-electrons can occupy all discrete levels, provided that the Debye screening length $\lambda_{\mathrm{D}}$ is sufficiently large. In this case, the Fermi sum $\sum_{n = 1}^{\infty} B_n(Z + 2)$ is enhanced by $3 \textrm{--} 5$ orders of magnitude and some of the decay-rate ratios $\Gamma^{0 \nu (2 \nu) \mathrm{EP} \beta}/\Gamma^{0 \nu (2 \nu) \beta \beta}$ exceed unity. The effect can be interpreted as follows: the sum $\sum_{n = n_{\mathrm{min}}}^{\infty} R_{n0}^2(0) \sim Z/(n_{\mathrm{min}} - n_a)^2$ from Eq.~(\ref{eq:q}) is replaced due to the full ionization by its hydrogen-like analog $\sum_{n = 1}^{\infty} R_{n0}^2(0) \sim Z^3$. For the parent isotope $\ce{_{34}^{82}Se}$ with $n_{\mathrm{min}} = 5$ (for the $n\mathrm{s}_{1/2}$ states) and $n_a = (3Z/2)^{1/3}$, the enhancement factor can be estimated to give $\approx 2 \times 10^3$, and it increases with $Z$. The $0 \nu \mathrm{EP} \beta$ decay channel becomes the only possible one for the fully ionized atoms of $\ce{^{98}Mo}$ and $\ce{^{146}Nd}$, in addition to $\ce{^{80}Se}$, $\ce{^{114}Cd}$, $\ce{^{122}Sn}$, $\ce{^{134}Xe}$ and the rest of double-$\beta$-decaying isotopes starting from $\ce{^{170}Er}$ in the case of the $2 \nu \mathrm{EP} \beta$ decay.

In plasma conditions, there is a shift and broadening of the atomic levels which affect the bound-state decay rates \cite{Sob72}. In an extreme case when the Debye screening length $\lambda_{\mathrm{D}}$ decreases below the Bohr radius $a_0$, the discrete levels of atoms are pushed to the continuum and, as a result, the bound states cease to exist. This phenomenon is known as the Mott transition \cite{Kra86}. In the cores of the Sun and Sun-like stars where $\lambda_{\mathrm{D}} \lesssim a_0$, the discrete levels of hydrogen are nonexistent. A similar situation occurs in the inner layers of white dwarfs. In the radiative zone of the Sun, e.g., where $\lambda_{\mathrm{D}} = (0.7 \textrm{--} 4) \, a_0$, the lowest discrete levels of hydrogen become a discrete part of the spectrum but remain vacant because of the ionization. The bound-state double-$\beta$ decays can thus occur in the outer layers of stars where the screening length is sufficiently large.

\section{Conclusion}
\label{sec:c}
In this paper, we studied the bound-state two-neutrino and neutrinoless double-$\beta$ decays. The corresponding phase-space factors were calculated in the framework of the $\mathrm{V} - \mathrm{A}$ weak-interaction theory including the mixing of Majorana neutrinos. The continuum wave functions of the $\beta$-electrons were approximated by the solutions to the Dirac equation in the Coulomb potential of the daughter nucleus, while the relativistic bound-electron wave functions, which are sensitive to the electron-shell screening effects, were computed via the multiconfiguration Dirac--Hartree--Fock package \textsc{Grasp2K}. The ratios between the decay rates of the bound-state and continuum-state double-$\beta$ decays, which are independent of the nuclear matrix elements and the effective Majorana neutrino mass, are maximal for the isotopes with lowest $Q$ values.

The bound-state double-$\beta$ decays were found to be several orders of magnitude less probable than the continuum-state double-$\beta$ decays. The bound-state neutrinoless channel is therefore not very suitable in the searches for lepton number violation. In contrast, the sensitivity of the modern $0 \nu \beta \beta$-decay experiments is already sufficient to observe the $2 \nu \mathrm{EP} \beta$ decay mode. We propose to set experimental limits on the $0 \nu \mathrm{EP} \beta$ peak and study the $2 \nu \mathrm{EP} \beta$ one-electron spectra in the tracking-and-calorimetry double-$\beta$-decay experiment NEMO-3 and its next-generation successor SuperNEMO, and examine the two-electron spectra in the calorimetric experiments CUORE \cite{Ald18}, EXO-200 \cite{Alb18} and GERDA \cite{Ago18}, as well as their upcoming tonne-scale upgrades.

Since under the standard conditions for pressure and temperature most of the double-$\beta$-decaying isotopes are solids, it would be desirable to generalize the proposed formalism to the scenario in which the electron shells belong to atoms embedded in a crystal lattice.

\begin{acknowledgments}
The authors are grateful to R.~Dvornick\'{y} for useful discussions of techniques adopted in the atomic-structure calculations. Special thanks go to the participants of the MEDEX'17 meeting in Prague (May 29--June 2, 2017) for their helpful comments and suggestions. This work was supported by the VEGA Grant Agency of the Slovak Republic under Contract No.~1/0922/16, the Slovak Research and Development Agency under Contract No.~APVV-14-0524, the Ministry of Education, Youth and Sports of the Czech Republic under the ERDF Grant No.~CZ.02.1.01/0.0/0.0/16\_013/0001733, the Grant of the Plenipotentiary Representative of the Czech Republic in JINR under Contract No.~192 from 05/04/2018, the RFBR Grant No.~18-02-00733, and the Grant of the Heisenberg--Landau Program under Contract No.~HLP-2015-18.
\end{acknowledgments}

\nocite{*}
\bibliography{0n2nEPb}
\end{document}